# Temperature dependent dynamics in water-ethanol liquid mixtures

*Szilvia Pothoczki[1]. László Pusztai[1,2] and Imre Bakó[3]*

[1]Wigner Research Centre for Physics, Hungarian Academy of Sciences, H-1121 Budapest, Konkoly Thege M. út 29-33., Hungary

[2]International Research Organization for Advanced Science and Technology (IROAST), Kumamoto University, 2-39-1 Kurokami, Chuo-ku, Kumamoto 860-8555, Japan

[3]Research Centre for Natural Sciences, Hungarian Academy of Sciences, H-1117 Budapest, Magyar tudósok körútja 2., Hungary

## Abstract

Temperature dependent hydrogen bond energetics and dynamical features, such as the diffusion coefficient and reorientational times, have been determined for ethanol-water mixtures with 10, 20 and 30 mol % of ethanol. Concerning pairwise interaction energies between molecules, it is found that water-water interactions become stronger, while ethanol-ethanol ones become significantly weaker in the mixtures than the corresponding values characteristic to the pure substances. Concerning the diffusion processes, for all concentrations the activation barrier of water and ethanol molecule become very similar to each other. Reorientation motions of water and ethanol become slower as ethanol concentration is increasing. Characteristic reorientational times of water in the mixtures are substantially longer than these values in the pure substance. On the other hand, this change for ethanol is only moderate. The reorientation motions of water (especially the ones related to the H-bonded interaction) become very similar for those of ethanol in the mixtures.

## 1. Introduction

Aqueous binary mixtures are of great importance in chemistry and biology. Mixtures of water and alcohols are one of the simplest systems in which there is a competition between the hydrophobic and hydrophilic (hydrogen-bonding) interaction in defining the properties of the system. It is also well-known that the thermodynamic and transport properties (diffusion coefficient, reorientation correlation time) of their mixtures show an anomalous behavior [1-15]. The anomalies of liquid water are more pronounced in the low temperature regime [16-19]. In most cases, analogous non-ideal behaviour is more pronounced (showing minima or maxima) in the low alcohol concentration region. Despite the large efforts in order to construct a well-defined atomistic picture [20-27] and a molecular-scale understanding of the behavior, no single, widely accepted model exists for these liquid mixtures.

The perturbation of the hydrogen-bond (HB) network is thought to be one of the reasons behind these anomalous properties. One of the first explanations was proposed by Frank and Ewans [1], suggesting the formation of an 'iceberg' (clathrate) like hydration shell around the hydrophobic moiety of the alcohol molecule. In this shell the strength of H-bond would be significantly stronger than in bulk water. There are a quite a lot of theoretical and experimental evidence for [1,5,7,8,14] and against [12,15] this model in the literature  Furthermore, it is known that near the hydrophobic surface the translational and orientational motions of water molecules are retarded. Some of the authors connected the activation energy of the reorientational motion to the energy of H-bond breaking [35,43,50].

Quite recently we studied the structural changes in methanol and ethanol-water mixtures as a function of temperature in the water rich region [28,29]. In these work we focused mainly on the changing properties of cyclic entities. We found in both systems that the number of hydrogen bonded rings has increased with lowering the temperature. However, for ethanol-water mixtures the dominance of not the six-, but of the five-fold rings could be observed.

One of the main goals of the present work is to describe changes of the interaction energy between the constituent molecules. To this end, we explore more accurately the energetics of the interactions around water and ethanol molecules in 2 dimensions (OO distance and energy). We also study that how some important dynamical properties (diffusion constant, reorientation correlation times) change as a function of the temperature in these mixtures.

## 2. Computational details

All the molecular dynamics simulations were performed by the GROMACS simulation package [30] (version 5.1.1), using the leap-frog algorithm for integrating Newton's equations of motion, with a time step dt=2 fs. Essential simulation parameters of the models (box lengths, number of ethanol and water molecule) are listed in Table 1.

Table 1 Temperatures, box lengths, number densities and bulk densities of the simulated systems.

| $x_e$ | T (K) | L (nm) | number density (atom/$\text{Å}^3$) | density (g/cm$^3$) | number of ethanol molecules | number of water molecules |
|---|---|---|---|---|---|---|
| 0.10 | 298 | 4.6900 | 0.1173 | 1.126 | 336 | 3024 |
| 0.10 | 268 | 4.8892 | 0.1035 | 0.994 | 336 | 3024 |
| 0.10 | 258 | 4.8850 | 0.1038 | 0.997 | 336 | 3024 |
| 0.10 | 253 | 4.8802 | 0.1041 | 0.999 | 336 | 3024 |
| 0.20 | 298 | 4.9500 | 0.0997 | 0.932 | 576 | 2304 |
| 0.20 | 268 | 4.8889 | 0.1035 | 0.967 | 576 | 2304 |
| 0.20 | 258 | 4.8752 | 0.1044 | 0.975 | 576 | 2304 |
| 0.20 | 253 | 4.8752 | 0.1044 | 0.975 | 576 | 2304 |
| 0.20 | 243 | 4.8560 | 0.1056 | 0.987 | 576 | 2304 |
| 0.20 | 233 | 4.8489 | 0.1061 | 0.991 | 576 | 2304 |
| 0.30 | 298 | 5.1900 | 0.0865 | 0.791 | 756 | 1764 |
| 0.30 | 268 | 4.8903 | 0.1034 | 0.946 | 756 | 1764 |
| 0.30 | 253 | 4.8683 | 0.1048 | 0.959 | 756 | 1764 |
| 0.30 | 238 | 4.8405 | 0.1066 | 0.975 | 756 | 1764 |

All simulations used the 'all atom type' OPLS-AA potential [31] for ethanol and the SPC/E [32] model for water. The cut-off radius for non-bonded interactions was set to 1.1 nm. All the simulations have been conducted with N>1000 molecules. In an earlier study, Gereben et al. showed [33] that such a system size may be used to study the dynamical properties of water.

Initially, an energy minimisation procedure was performed for each composition at room temperature, using the steepest descent method. This was followed by a 5 ns equilibration run in the NPT ensemble; the temperature and pressure were controlled by a Berendsen thermostat and barostat [34], with temperature coupling time constants set to 0.1 ps and 0.5 ps, respectively. Following this long equilibration procedure, using an additional 1 ns production runs in NVT ensemble were carried out, in which particle the obtained configurations were saved in every 10 steps for additional statistical analyses.

The diffusion coefficient (D) was estimated using the Einstein-Smoluchowski relation, from the mean squared displacements of the centres of mass of water and ethanol molecules:

$$D = \lim_{t \to \infty} \frac{1}{6Nt} \left\langle \sum_{i=1}^{N} \left( r_i(t) - r_i(0) \right)^2 \right\rangle \tag{1}$$

where $r_i(t)$ and $r_i(0)$ are the positions of the centres of mass of water or ethanol molecules at time t and 0, respectively, and the <...> denotes an ensemble average. The effect of using every x-th saved configuration (x=1,5,20) during the MSD calculation was negligible, as it has already been shown by Gereben et al. [33] for the SPC/E water model.

Reorientational dynamics have been characterized by the autocorrelation functions:

$$C_l(t) = \langle P_l(\underline{e}(t) \cdot \underline{e}(0) \rangle \qquad (2)$$

where $\underline{e}(t)$ is the unit vector along a well-defined molecular axis (O-H vector, perpendicular to the HOH water molecular plane, or $C_1C_2O$ ethanol plane) and $P_l$ is the l-th Legendre polynomial. The characteristic decay time $C_2(\underline{OH}(t))$ is measurable using NMR experiments [35]. The decay time of these autocorrelation functions, $<\tau>$, is estimated by computing the integral of $C_l(t)$ with respect to time, that is:

$$\tau = \int_0^\infty C_l(t)dt \qquad (3)$$

## 3. Results and Discussions

### 3.1 Energy distributions

A deeper analysis of the composition and temperature dependence of the strength of intermolecular associations of water and ethanol molecules can be performed by studying the pair energy (Coulomb + Lennard-Jones terms) distributions. The computed pair energy distributions for pure ethanol and SPC/E water are shown in Fig.1.

Fig.1. Pair energy distributions for pure liquid ethanol and water as a function of temperature

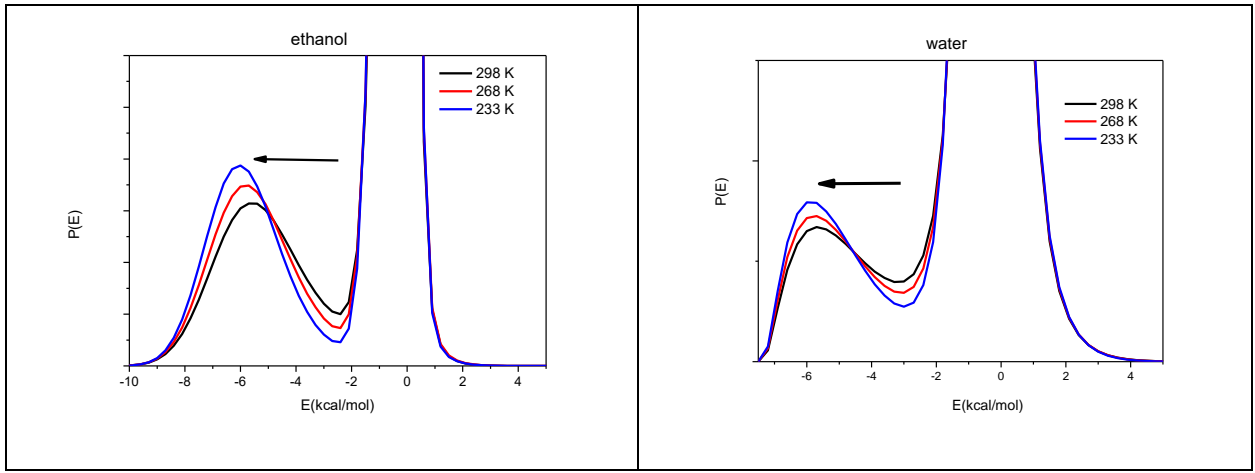

The pair energy distribution of H-bonded liquids has a characteristic shape, with (1) a spike near 0.0 kcal/mol that represents the interaction with distant molecules in the bulk, and (2) a low energy band for hydrogen bonded neighbors (following the first well defined minimum). The distribution of pair energies for water-water ('wa-wa') and ethanol-ethanol ('et-et') interactions in pure water and ethanol exhibits peaks at negative values of $E_{ij}$ at $-5.4$–$6.0$ kcal/mol, where the position of maxima decreases with decreasing temperature. We were able to identify a minimum after the first maximum for water and ethanol at 3.0 and 2.4 kcal/mol, respectively. The average pair interaction energy of ethanol molecules that corresponds to the strongly interacting (H-bonded) dimers changes from -5.4 kcal/mol to -5.8 kcal/mol as the temperature is decreased from 298 K to 233 K. The same quantity for water changes from -5.03 kcal/mol to -5.23 kcal/mol over the same temperature range.

In order to better understand these changes we have calculated the O-O distance-energy distribution for pure liquid water and ethanol; these data are presented in Fig. 2. This distribution allows us to find more precise energetic criteria for H-bond definition, applicable during studies liquid mixtures of ethanol and water. This is demonstrated in Fig. 3: a threshold at about -3.0 kcal/mol may be set for a proper H-bond definition. In pure liquid water, this is also an accepted value for H-bond definition [36].

Fig.2. shows that there is a significant change for H-bonded ethanol and water dimers as the temperature is decreasing. Additionally, we can detect another well-defined change in these plots at around 3.6 Å and +1.0 kcal/mol, for both molecules (cf. Fig.2.). At lower temperatures this region is getting more populated and more sharply defined.

Fig. 2. Distance-pair energy distributions for pure liquid ethanol and water at 298 and 233 K. The positions where significant changes may be detected are denoted by red arrows.

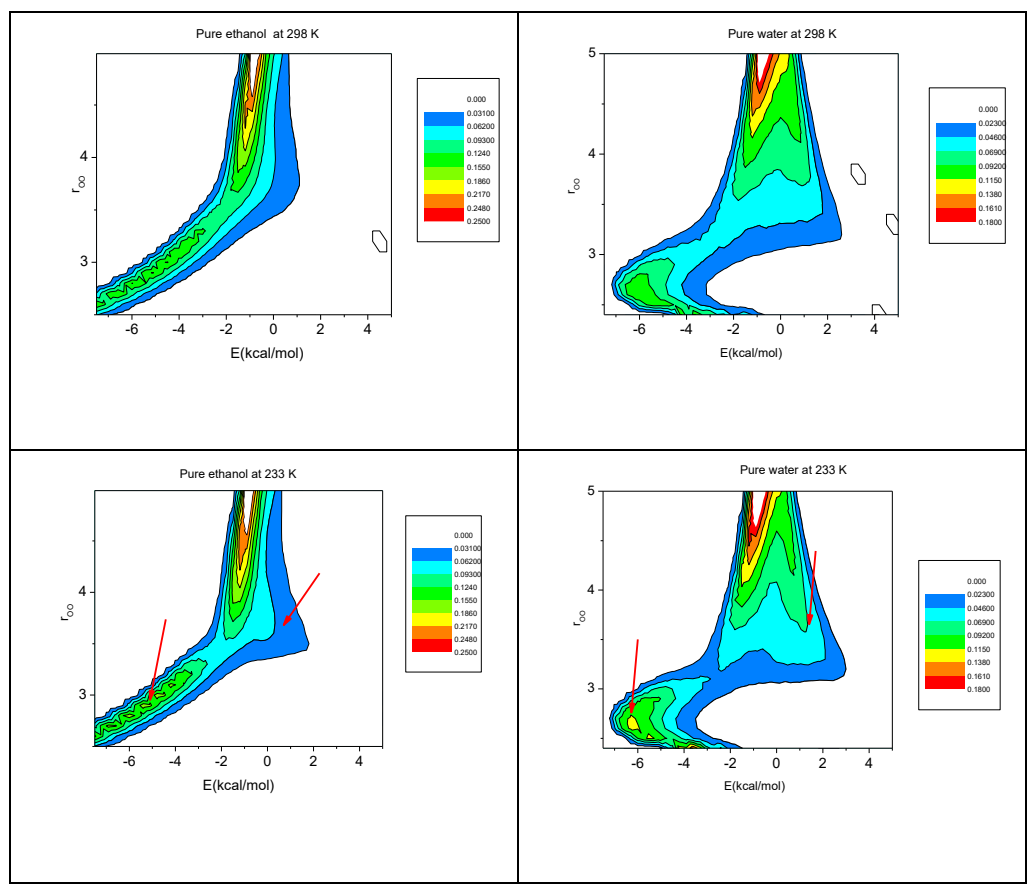

Fig. 3. Average pair energies (left) and their distributions (for one composition only) for ethanol-water mixtures, as a function of temperature.

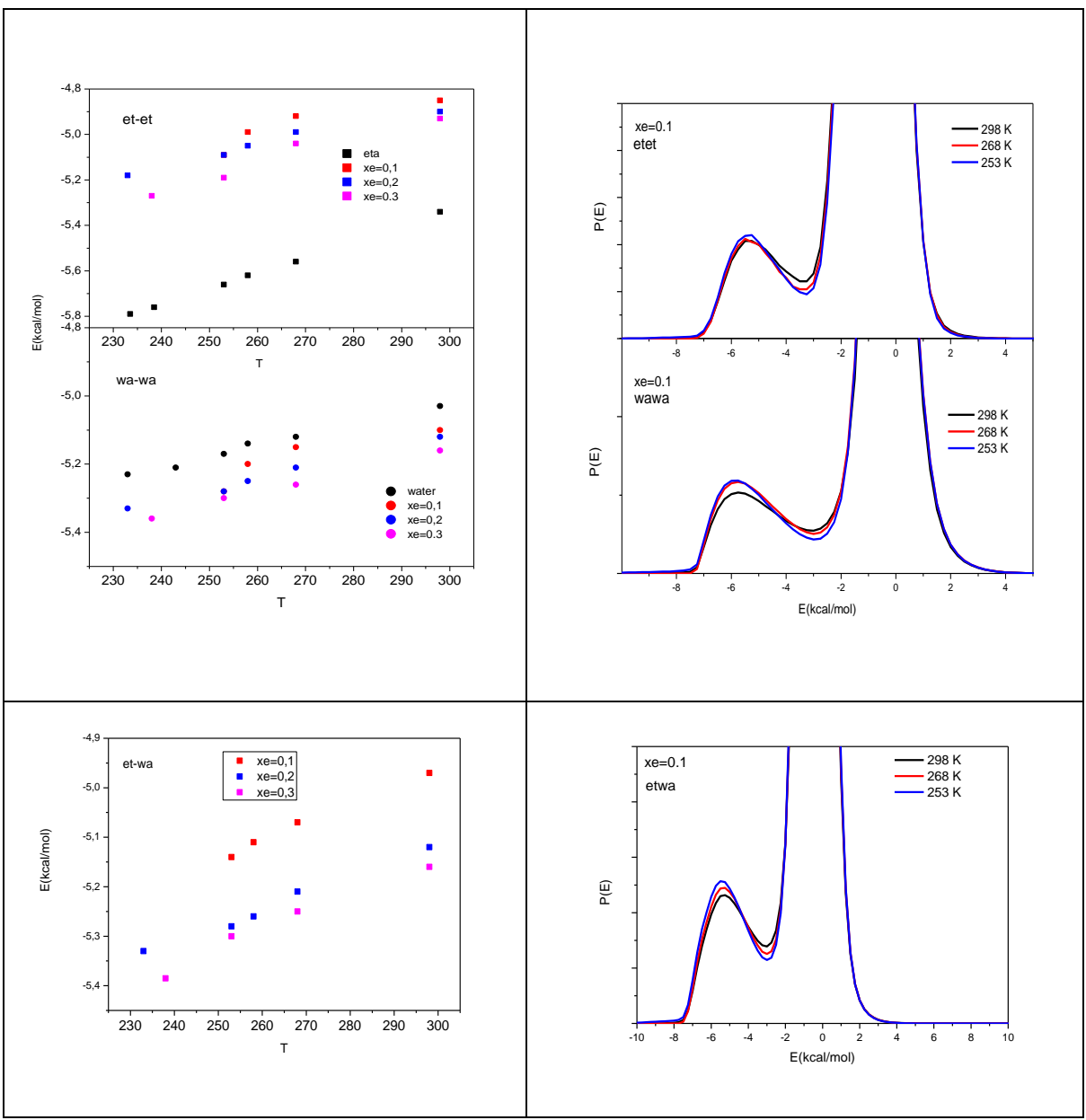

In addition, we have calculated the average pair interaction energy for water and ethanol as a function of temperature (see Fig. 3). Clearly, the pair interaction energy is becoming more negative on decreasing temperature. Note that water-water interactions become stronger, while ethanol-ethanol ones become significantly weaker in the mixture than the corresponding values characteristic to the pure substances. On the other hand, variations that may be detected in terms of the 2D (distance-energy) distributions (see Fig. 4) in the mixture are similar to what we have already detected in pure water and ethanol (see above).

Fig. 4. Distance-pair energy distributions for the ethanol-water mixture with 20 mol % of ethanol at 298 and 233 K. The positions where significant changes may be detected are denoted by red arrows.

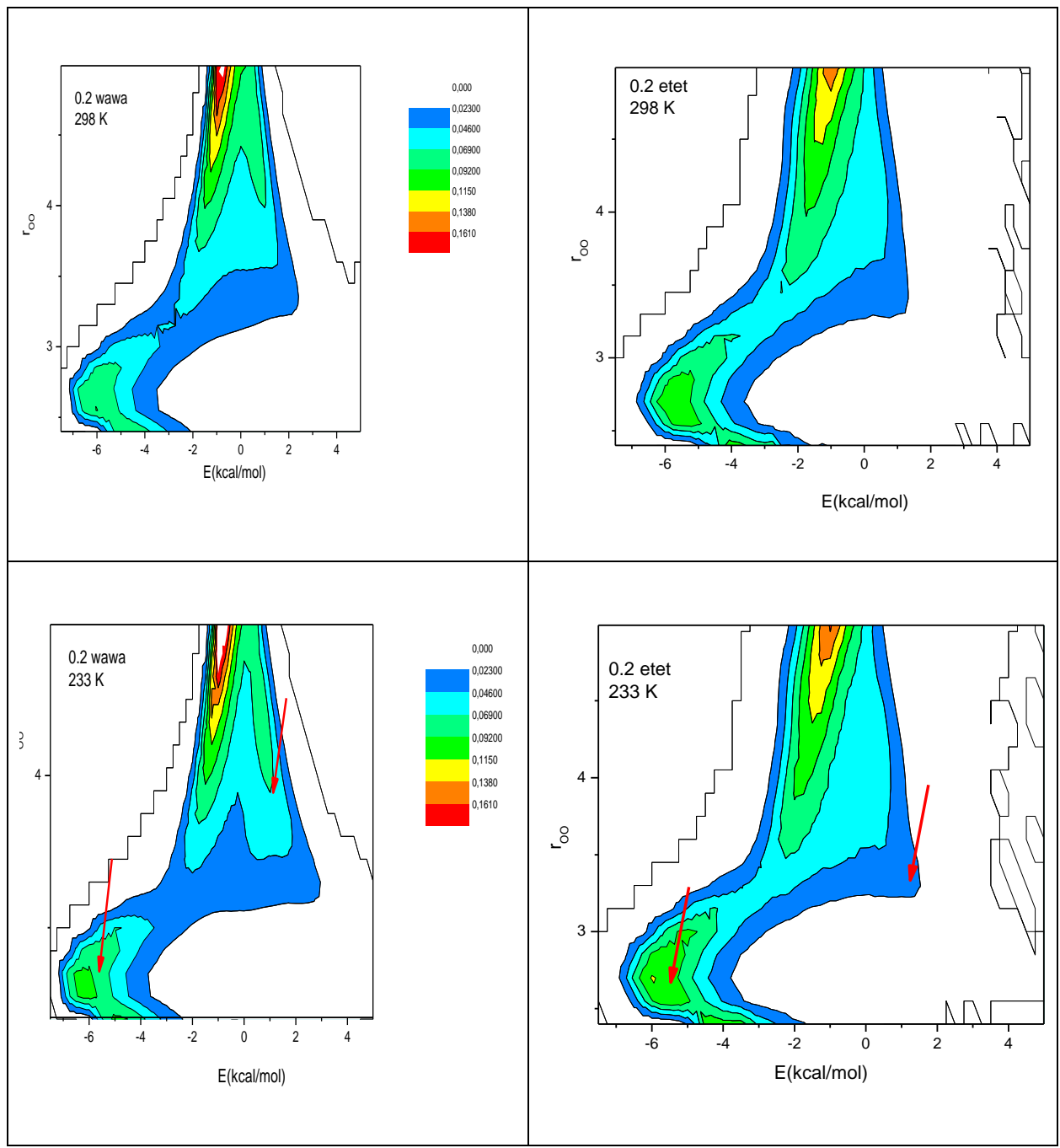

*3.2 Diffusion coefficients*

Mean squared displacements (MSD) of centers of mass as a function of time are used here to calculate the self-diffusion coefficient by Einstein's method. In Fig. 5, MSD plots for pure ethanol (a) and water (b) are shown at different temperatures. It is clear from this figure that the MSD-s of water and ethanol molecules show diffusive behavior over the timescale of our calculations even at the lowest temperature. The MSD is becoming steeper with increasing temperature; this shows that the rate of diffusion is increasing with increasing temperature. In order to validate our computational procedure, we compare simulated data for liquid water and ethanol with data from literature [42,43,44], using the same potential model. The statistical accuracy of the calculated diffusion coefficient is about 1-2 %.

Fig 5. Mean square displacements for simulated pure ethanol (a) and water (b) at different temperatures.

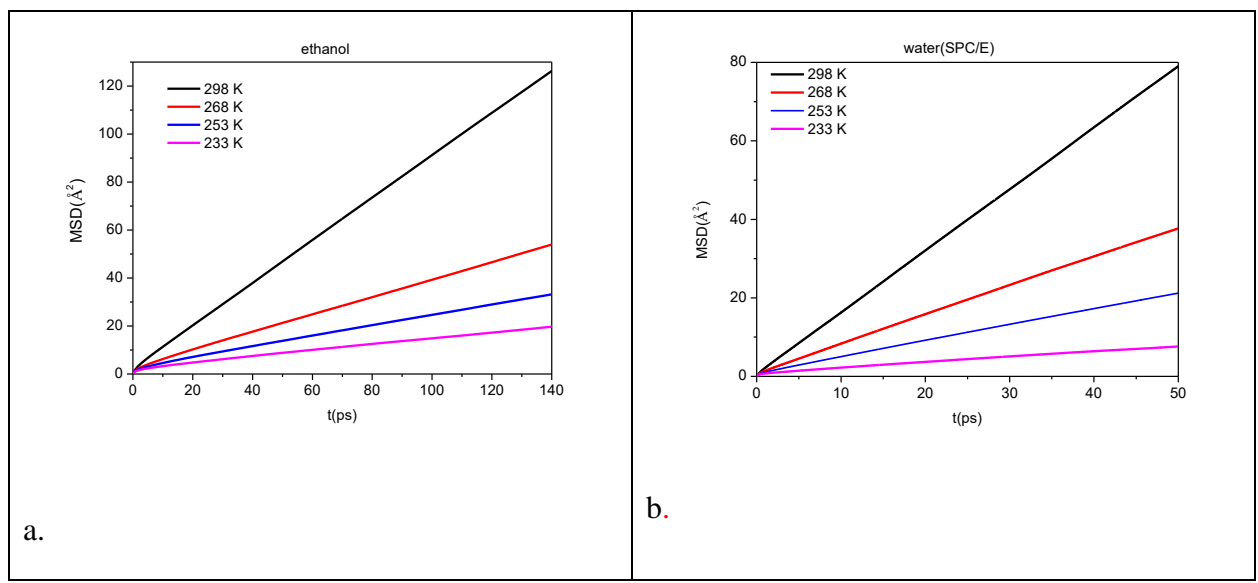

Fig. 6. Experimental and simulated diffusion coefficients for liquid ethanol at different temperatures (simulation by Hasse (42), experiment-1 [38,39]. experiment-2 [40, 41]). Data are shown as Arrhenius-type plots.

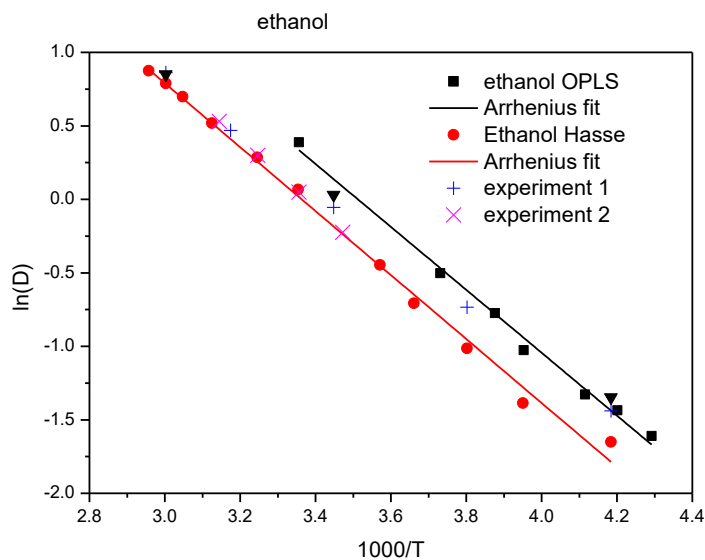

In Figures 6 and 7 we plot ln(D) for pure ethanol and water obtained from our MD simulations and from experiments [36-42] as a function of the inverse temperature. In the case of water our data in the investigated temperature range agree well with results of Galamba [43]. The difference from experimental data is about 5-6 % at room temperature and about 8-10 % at low temperature. Here we would like to remark that the experimental uncertainty of the self-diffusion coefficient is about 10%. It is clear from this figure, however, that the temperature dependence of experimental self-diffusion coefficient does not have an Arrhenius like behavior, especially not at low temperature. This non-Arrhenius behavior could not be reproduced by MD simulations.

Fig. 7. Experimental and simulated diffusion coefficients for liquid water at different temperatures (simulation by Galamba [43], experiment-1 [36]. experiment-2 [37]). Data are presented as Arrhenius-type plots.

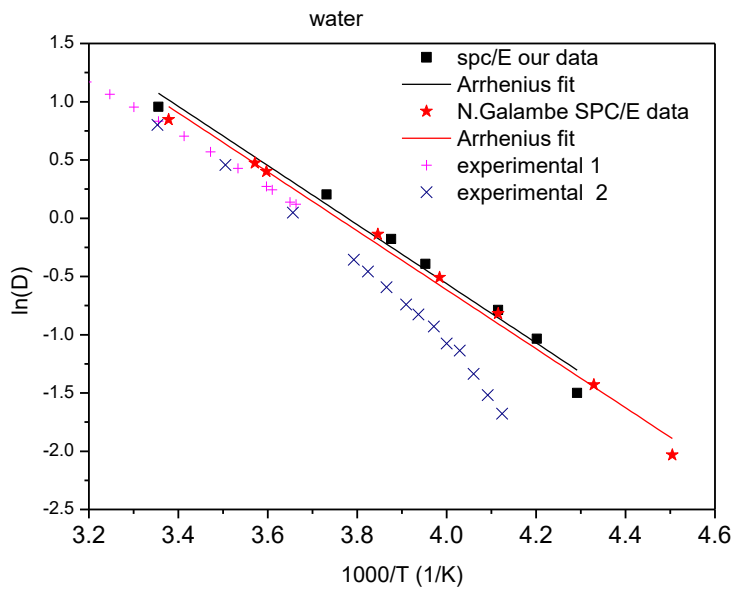

The temperature dependence of our results for $D_w$ and $D_e$ over the temperature range 298 to 253 K can be reasonably well described by Arrhenius plots, as shown in Figures 6 and 7. Values of the activation energy, which can be used as a direct measure of the temperature dependence of the self-diffusion coefficient, are presented in Table 3. For pure water, it is about 21.1 kJ/mol from both our and Galamba's simulation data [43]. The activation energies reported for the experimental D of water over the range of temperatures 273 to 323 K [38,39] are around 21.5 kJ/mol.

Table 2. Calculated diffusion coefficients of the components as a function temperature (in bracket experimental values ([48,49 ])

| $x_{ethanol}$ | T(K) | D (ethanol,$10^{-9}$ m$^2$/s) | D (water,$10^{-9}$ m$^2$/s) |
|---|---|---|---|
| 0.1 | 298 | 1.03 (0.65,0.718) | 1.70 (1.0,1.26) |
| | 268 | 0.40 | 0.67 |
| | 258 | 0.23 | 0.40 |
| | 253 | 0.18 | 0.34 |
| 0.2 | 298 | 0.81(0.5,0.618) | 1.26(0.8,0.99) |
| | 268 | 0.35 | 0.52 |
| | 258 | 0.21 | 0.30 |
| | 253 | 0.15 | 0.23 |
| | 243 | 0.09 | 0.13 |
| | 233 | 0.04 | 0.06 |
| 0.3 | 298 | 0.83(0.55,0.623) | 1.17(0.75,0.91) |
| | 268 | 0.34 | 0.45 |
| | 253 | 0.15 | 0.21 |
| | 238 | 0.07 | 0.09 |

The present calculated values are in agreement with other simulations using the OPLS-AA model for ethanol [22,26,44-47] that deviate from the experimental ones by an error margin of approximately 20-30 % above room temperature. With a small modification of this model, or using a slightly modified united atom model, the agreement becomes significantly better at room temperature. At low temperature the agreement between simulated and experimental results is also much improved: the calculated activation energy is about 17.9 kJ/mol and 17.4 kJ/mol for simulated and experimental data [40-43], respectively. The activation energy of the diffusion process in liquid ethanol is smaller than that in water.

Calculated MSD for water and ethanol in the $x_e$=0.2 mixture at different temperatures are shown in Fig. 8. It is clear that even at the lowest temperature (233 K) it was possible to calculate properly the diffusion constant using Eq. 2.

Fig. 8. Mean square displacements of ethanol (OPLS) and water (SPC/E) molecules at different temperatures in the mixture with 20 mol % ethanol.

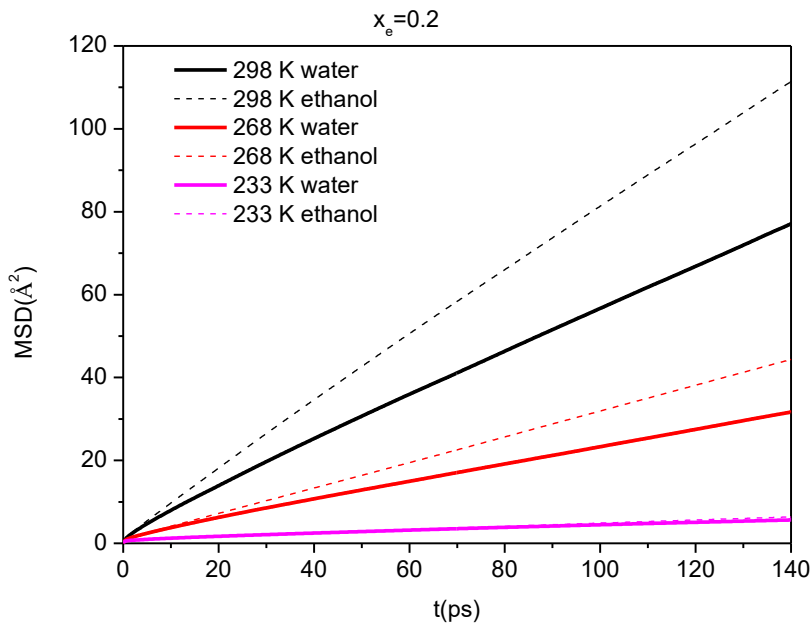

Calculated self-diffusion coefficients as a function of temperature are presented in Table 2. There seems to be a rapid decrease in the self-diffusion coefficients of both water and ethanol molecules in the water rich region of the mixtures from experimental data. This indicates that there is a well-defined change between the strengths of interactions of water and ethanol molecules. The behavior of our simulation data agrees qualitatively with the experimental finding, but quantitatively our data significantly overestimate the experimental results (by ca.20-40%) [48-50].

Calculated activation barriers for water and ethanol molecules are presented in Table 3. It appears that for all concentrations, the activation barriers of water and ethanol molecules become very similar.

Table 3 Activation energies for D as clculated from MD simulations

|  | Water | Ethanol |
|---|---|---|
| Water | 21.1±0.8 |  |
| $X_e$=0.1 | 22.6±0.6 | 24.2±0.7 |
| $X_e$=0.2 | 26.8±0.9 | 26.6±0.8 |
| $X_e$=0.3 | 25.2±0.8 | 24.7±0.7 |
| Ethanol |  | 17.9±0.6 |

*3.3 Reorientational correlations*

The reorientational relaxation dynamics of liquid water and ethanol have been investigated recently by theoretical and experimental methods alike [35,43,49-53] Here we characterize the reorientational dynamics of ethanol and water molecules by autocorrelation functions of the OH-groups and of vectors normal to the HOH (water) and CCO (ethanol) plane, as described in the 'Methods' section. From the $C_2(OH)(t)$ function we can calculate a characteristic time which is directly related to the experimentally available reorientation time from NMR [49-52]. The calculated reorientational times, together with the same data from other simulations, and also experiments, for liquid ethanol (a) and water (b) are presented in Fig. 9.

Fig. 9. Reorientational correlation times for pure ethanol (OPLS) and water (SPC/E) as a function of 1000/T. Experiment on $CD_3CD_2OH$: [51,52]; experiment on $CH_3CH_2OD$ [53]; experiment-1 on water: [49]; experiment-2 on water: [50,52]; simulation on water by Galamba: [43]

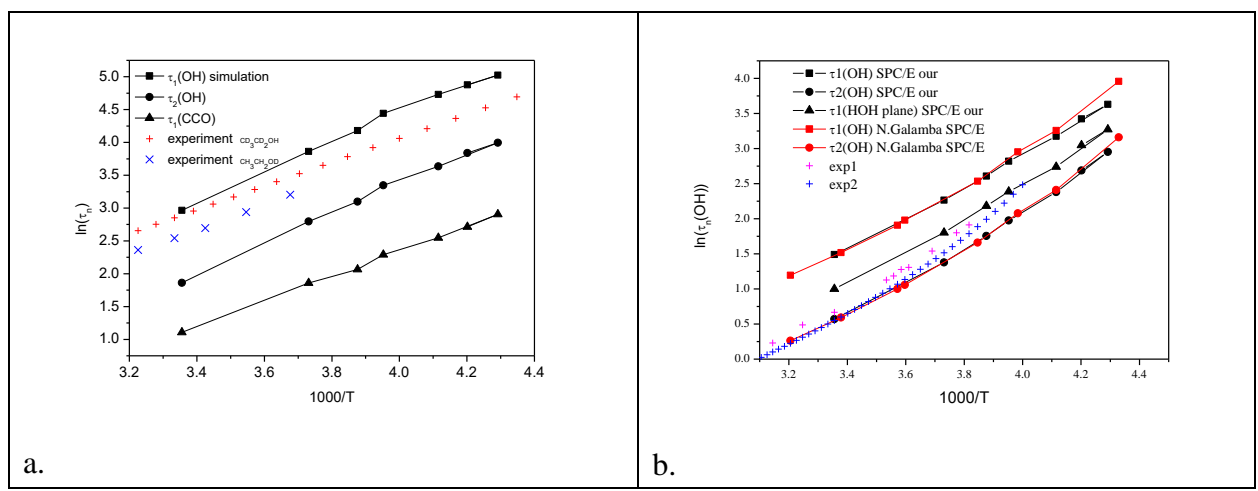

It is clear that we can reproduce the $\tau_2$ reorientational correlation times form an earlier simulation using the SPC/E water model over a broader temperature range reasonable well. Deviations from experiments over this range is 10-20 %. (The difference is larger at low temperature.) The reorientational correlation time $\tau_1$ of the HOH plane is significantly shorter than the same quantity for the OH unit vector ($\tau_{1OH}/\tau_{1HOH}$ is in the range of 3-6), which strongly suggest the existence of a well-defined rotational anisotropy in liquid water; this is in good agreement with the experimental evidence. The calculated activation energy for $\tau_2$ is about 21.2 ($\pm$0.8) kJ/mol from simulation and 19.6 ($\pm$0.4) kJ/mol from experimental data.

Calculated reorientation times for liquid ethanol, together with the available experimental data for liquid $CD_3CD_2OH$ and $CH_3CH_2OD$, as a function of 1000/T are presented in Fig. 9a. The calculated activation energies from the experimental and MD results are 15.1kJ/mol, 16.8 kJ/mol, and 18.9 KJ/mol for $CD_3CD_2OH$ [51,52], $CH_3CH_2OD$ [53], and for simulation data, respectively. The $\tau_2$ value from simulation is significantly smaller (about ½-th or 1/3-th) than the corresponding experimental values. There are at least two different reasons for this behavior: (a) problems with the potential: OPLS-AA significantly underestimated this quantity, and (b) the NMR experiments were performed on liquid $CD_3CD_2OH$ and $CH_3CH_2OD$, but not on liquid $CH_3CH_2OH$.

The average integrated decay times, $\tau_1$, for the OH bond and for the vector normal to the CCO plane are significantly different in the investigated temperature range, showing well-defined orientation anisotropy for liquid ethanol, which is significantly larger than in the case of liquid water. The slower reorientational dynamics in liquid ethanol compared to water can be explained, as stated by Vartia et all. [35], by the so-called 'extended jump' model.

Characteristic reorientation times are presented in Figs. 10 and 11 as a function of temperature for the investigated mixtures and for the pure liquids. It can be concluded from these figures that the reorientation motions of water and ethanol molecules become slower as the ethanol concentration is increasing. The $\tau_1$, $\tau_2$ characteristic times of water is substantially larger than these values in pure substance. On the other hand, this change for ethanol is only moderate. The reorientation motions of water molecules (especially the ones related to the H-bonded interaction) become very similar for those of ethanol. We can prove this statement by Fig. 12, where the ratios of the corresponding decay times of the two molecules are presented as a function of temperature for the $x_e$=0.2 mixture and for the pure liquids . The calculated ratio for the pure substance is dramatically different from 1, and at the same time, it is very close to unity for the mixtures in the investigated concentration range.

Fig. 10. Reorientation correlation times of ethanol molecules in the mixtures and in pure liquid at different temperature.

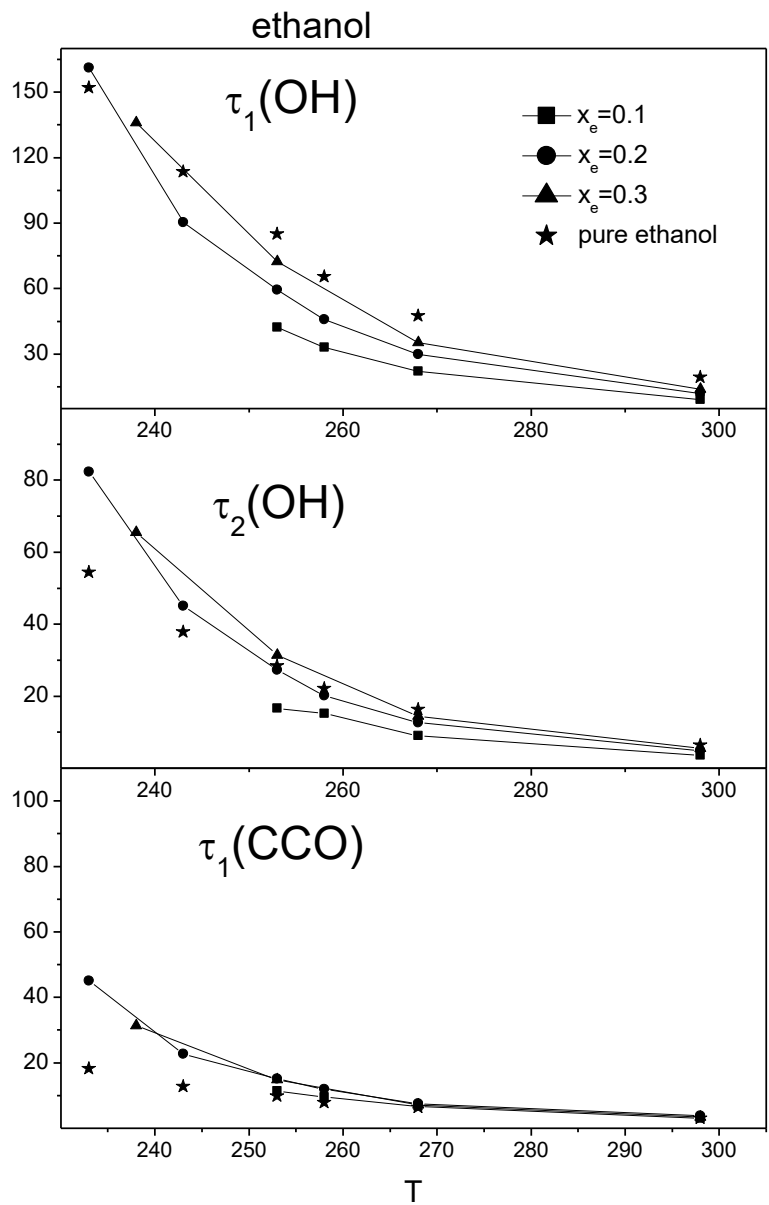

Fig. 11 Reorientation correlation times of water molecules in the mixtures and in pure liquid at different temperature.

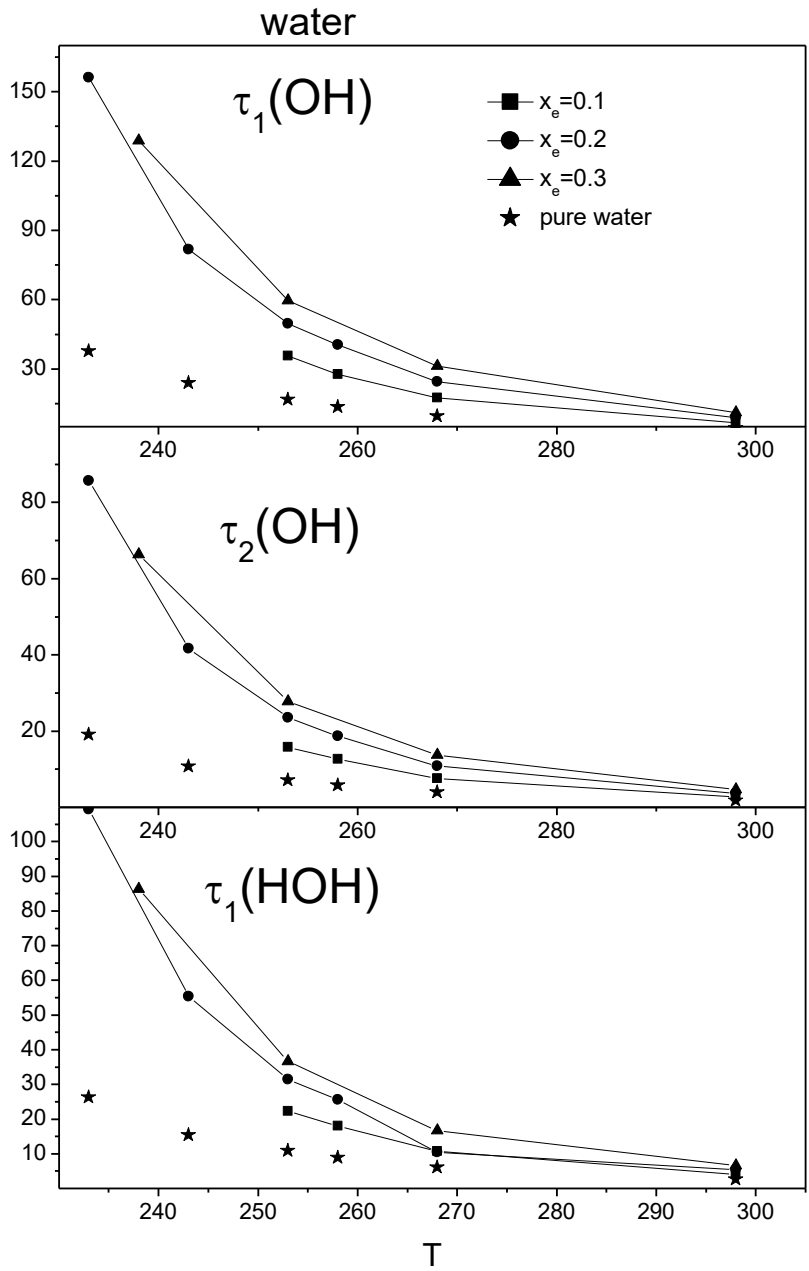

Fig. 12 Ratios of the ethanol and water reorientation times in pure liquid and in the mixture ($x_e$=0.2) at different temperature.

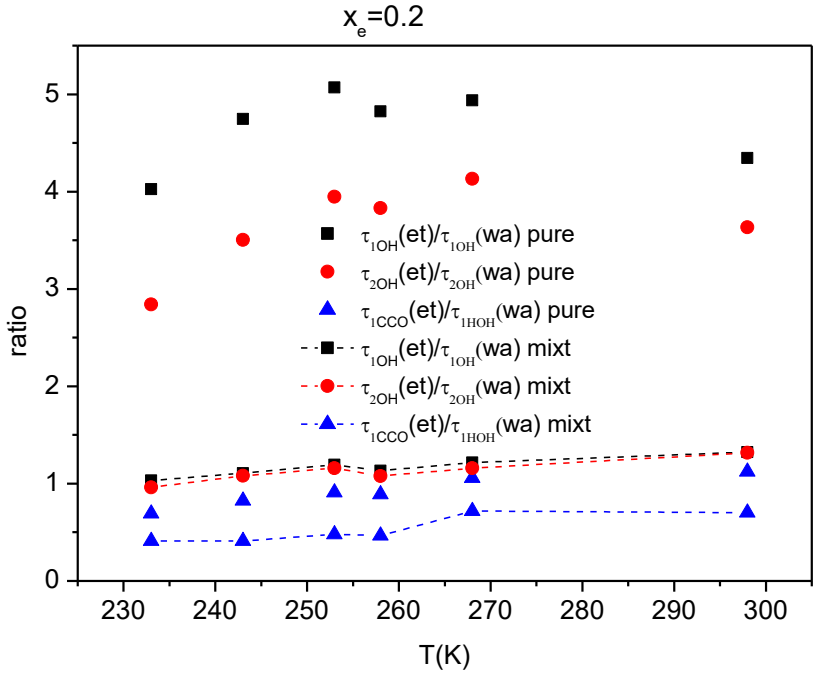

Calculated activation barriers of the reorientation motions for water and ethanol molecules are presented in Table 4. It is clear from these data that for all concentrations the activation barrier of both water and ethanol have a well-defined maxima at the composition of $x_e$=0.2. The activation barrier for reorienting water molecules in the mixture is larger than the corresponding value for ethanol molecules.

Table 4 Calculated activation barriers for water and ethanol molecules. (Data are in kJ/mol.)

|  | water | | ethanol | |
| --- | --- | --- | --- | --- |
|  | τ1(HOH) | τ1(OH) | τ1(CCO) | τ1(OH) |
| water | 20.2±0.8 | 21.4±0.4 |  |  |
| $x_e$=0.1 | 23.4±0.9 | 22.9±0.5 | 21.0±0.8 | 18.2±0.4 |
| $x_e$=0.2 | 26.6±1.0 | 25.2±0.3 | 23.0±1.1 | 21.6±0.4 |
| $x_e$=0.3 | 25.2±0.8 | 23.9±0.4 | 22.5±1.2 | 20.9±0.3 |
| ethanol |  |  | 15.8±0.8 | 19.0±0.5 |

## 4. Conclusions

Detailed analyses of the pair energies, as well as diffusional and reorientational motions of the molecules, as a function of composition and temperature, are presented for ethanol-water liquid mixtures in the water-rich region.

Concerning pairwise interaction energies between molecules, water-water interactions become stronger, while ethanol-ethanol ones become significantly weaker in the mixtures than the corresponding values characteristic to the pure substances. Additionally, in pure liquids and also in the mixtures we detected a substantially change in the interstitially region (3.6 Å, +1 kcal/mol).

Concerning self-diffusion, mean squared displacements of water and ethanol molecules clearly show diffusive behavior over the timescale of our calculations even at the lowest temperature (close to the experimental freezing point). Calculated activation barriers for diffusive motions of water and ethanol molecules become very similar in the liquid mixtures.

Various reorientational times for water and ethanol molecules have been determined in the pure liquids, as well as in the mixtures. The reorientation motions of both water and ethanol molecules become slower as the ethanol concentration increases. The $\tau_1$, $\tau_2$ characteristic times of water are substantially larger in the mixtures than these values in the pure substance. The activation barrier for reorienting water molecules in the mixture is larger than the corresponding value for ethanol molecules.

## Acknowledgments

The authors are grateful to the National Research, Development and Innovation Office (NRDIO (NKFIH), Hungary) for financial support via grants Nos. SNN 116198 (Sz. P. and L. P.) and 124885 (I. B.).